\documentclass[prd,aps,showpacs,preprintnumbers,amssymb]{revtex4}
\usepackage{axodraw}
\usepackage{color}
\usepackage{epsf}

\def\e3p{$\eta \rightarrow 3 \pi$}

\begin{document}
\title{%
\hfill{\normalsize\vbox{%
\hbox{}
 }}\\
{About flavor, spin and color}}

\author{Renata Jora $^{\it \bf a}$~\footnote[2]{Email:
 rjora@ifae.es}}
 \affiliation{$^ {\bf \it a}$ Grup de Fisica Teorica and IFAE, Universitat Autonoma de Barcelona,
 E-08193 Bellaterra(Barcelona), Spain.}

\date{\today}

\begin{abstract}
Chiral symmetry breaking (restoration) for $SU(N)$ gauge theories is a topic of great interest and not yet fully explained.  We consider the phenomenon as a collective spin effect and determine its behavior in terms of the number of flavors, $N_f$.
\end{abstract}


\maketitle
\section{Introduction}

The phenomenon of chiral symmetry breaking together with its counterpart chiral symmetry restoration is of great importance for nonabelian gauge theories since it can shed light on important issues in QCD and it is crucial for technicolor ones. The subject has been treated in detail from various points of view in \cite{Chivukula}-\cite{Schechter}
using gap equation or an instanton type of potential. All work done agrees with the picture that the chiral symmetry is broken by the quark condensate and  that as $N_f$ approaches a critical value $N_f^c$ the chiral symmetry is restored.
The current understanding of the chiral symmetry breaking (restoration) treated as a function
of the number of flavors  $N_f$ and colors N requires the two loop beta function and the anomalous dimension of the fermion mass operator.

In this paper we will present a different point of view. Thus instead of relying on the beta function we will base our two methods on the one loop effective potential. This is viewed on two distinct cases
as a function of a spin dependent operator which we associate with the paramagnetic magnetization and a spin independent one corresponding to the diamagnetic magnetization. It turns out that for each situation the potential displays two minima which according to Landau theory should indicate phase transitions in the system. We associate one
phase transition with the breaking(restoration) of chiral symmetry while we indicate that the second one could correspond to the baryon symmetry breaking (restoration).
Another argument in favor of our approach is the following: the spin dependent part in the one loop potential
corresponds to the zero modes in the non-perturbative language while the spin independent one corresponds to the positive modes.  This separation is often very important in quantum field theories: one loop supersymmetric potential contains only zero modes; breaking of $U(1)_A$ is proven to be due to zero fermion modes in the instanton approach. Thus we are motivated to think that zero modes might play a fundamental role also in the chiral symmetry breaking.
Here we show that a phase transition occurs at $N_f=4N$. But standard studies of chiral phase transition state that this occurs very close
to $N_f\approx4N$. It is natural then to identify the two of them.

Sections II and III contain well known results about chiral symmetry breaking and one loop calculations in
 nonabelian gauge theories. In sections IV and V we present our approach based on one-loop effective potential
 in the background field perturbation theory.
Section VI is dedicated to conclusions.

\section{Chiral phase transition: review of the main approach}

It is well established that the phase transition in Yang Mills theories(for a recent most general review see for example \cite{Sannino})
 occurs at an infrared fixed point of the beta function according to :
\begin{eqnarray}
\beta({\alpha})=
-b\alpha(\mu)^2-c\alpha(\mu)^3-d\alpha(\mu)^4-....
\label{beta435267}
\end{eqnarray}
Here, (for an SU(N) gauge theory with $N_f$ flavors)
\begin{eqnarray}
&&b=\frac{1}{6\pi}(11N-2N_f)
\nonumber\\
&&c=\frac{1}{24\pi^2}\left(
34N^2-10NN_f-3\frac{N^2-1}{N}N_f\right)
\label{coef435678}
\end{eqnarray}

A phase transition occurs when $\alpha=\alpha_c$ or when the anomalous dimension of the fermion operator is equal to one($\gamma\approx1$)\cite{Georgi} which reads:
\begin{eqnarray}
\gamma=\frac{3}{4\pi}\frac{N^2-1}{N}\alpha_c=1
\label{cond56478}
\end{eqnarray}

Thus one can deduce the critical number of flavor by putting the condition that this critical value coincides with the infrared fixed point of the theory:
\begin{equation}
\alpha_{\ast}=-\frac{b}{c}
\label{con564789}
\end{equation}

The behavior of the theory at the phase transition can be read off from a gap equation or from an instanton type of potential etc (see \cite{Chivukula}-\cite{Schechter}).

\section{One loop calculations in non-abelian gauge theories}

In \cite{Nielsen} the vacuum energy for QCD with an arbitrary number of flavors is obtained using a
simple quantum mechanical approach:

\begin{equation}
E_{vac,QCD}=-\frac{(33-2N_f)Vg^2H^2}{96\pi^2}\ln\frac{\Lambda^2}{g|H|}
\label{vac42567}
\end{equation}

Here V is the volume and H is the external field and one can recognize easily in the coefficient in front of the expression the argument of the one loop beta function for QCD. From the calculations is evident that if we identify paramagnetism as a spin dependent response to an external magnetic field and diamagnetism as a spin independent one then the vacuum energy can be written as a sum of two terms, one paramagnetic and one diamagnetic:

\begin{eqnarray}
E_{vac, QCD}=E_{vac,para}+E_{vac,dia}
\label{form786950}
\end{eqnarray}

where( generalized to SU(N) non-abelian gauge theories),
\begin{eqnarray}
&&E_{vac,para}=-\frac{(4N-N_f)Vg^2H^2}{32\pi^2}\ln\frac{\Lambda^2}{g|H|}
\nonumber\\
&&E_{vac,dia}=-\frac{(-N+N_f)Vg^2H^2}{96\pi^2}\ln\frac{\Lambda^2}{g|H|}
\label{ne436567}
\end{eqnarray}

The full vacuum for QCD is "paramagnetic" if paramagnetism is viewed as a negative response to the action of an external magnetic field.

Alternatively it was pointed in \cite{Vainshtein} that this separation arises also in instanton calculations where the term proportional to $4N-N_f$ corresponds to zero frequency modes and that proportional to $(N_f-N)/3$ to positive frequency ones.

Observe that for $N=N_f$ the vacuum is purely paramagnetic (the vacuum energy is fully dependent on the spin interaction) and for $N=4N_f$ is purely diamagnetic (the vacuum energy does not contain spin interactions but only summation over the spin degrees of freedom).

Let us view things from the perspective of the "background field perturbation theory"(see for details \cite{Peskin})
which will be employed as a start-up in the present work.
Thus the effective action for an external field $A_{\mu}^a$ derived in the path integral formulation is:

\begin{eqnarray}
e^{i\Gamma[A])}=\int{\cal D}{\cal A}{\cal D}\Psi{\cal D}c
\exp[i\int d^4x [-\frac{1}{4g^2}(F^a_{\mu\nu})^2 +{\cal L}_{ct}]]
(\det \bigtriangleup_{G,1})^{1/2}(\det \bigtriangleup_{r,1/2})^{N_f/2}(\det \bigtriangleup_{G,0})^{+1}
\label{eff435267}
\end{eqnarray}

Here

 \begin{eqnarray}
 \Delta_{r,j}=-D^2+2(\frac{1}{2}F^a_{\rho\sigma}{\cal I}^{\rho\sigma})
 \label{term546738}
 \end{eqnarray}

 and r refers to representation while j refers to spin. One needs to compute determinants as:
\begin{eqnarray}
\det(A)=\exp[Tr(\log(A)]
\label{det342567}
\end{eqnarray}

We will write:
\begin{eqnarray}
\bigtriangleup_{r,j}=-\partial^2+\bigtriangleup^{1}+\bigtriangleup^{2}+\bigtriangleup^{j}
\label{not876}
\end{eqnarray}

We need to compute:
\begin{eqnarray}
\log(\det(\bigtriangleup_{r,j}))=
\log(\det[-\partial^2])+Tr[(-\partial^2)^{-1}(\bigtriangleup^1+\bigtriangleup^2+\bigtriangleup^3)]
\label{somformula64789}
\end{eqnarray}

and add all the relevant contributions.
Finally what one gets is a functional of the background field which essentially is the coupling constant at one loop.

\section{Chiral phase transition: a new perspective}

We need a little procedure to rewrite the one loop functional. But first let us envision the expression for the final result.
\begin{eqnarray}
V=e_2F^2+e_3F^3+e_4F^4+....
\label{potyru7465}
\end{eqnarray}

Here by $F^n$ we understand gauge invariants of the corresponding order that can be formed with the nonabelian  tensor. Since in the final result we can specify the nonabelian background gluon field, by the arguments given in the appendix we consider as relevant only the terms of even powers in the field tensor.

We plan to derive the correspondence between the number of flavors $N_f$ and the number of colors N at which the chiral symmetry restoration occurs using entirely the one loop perturbative potential. As explained in section II in the established literature this is done only with the aid of the two loop beta function. It is one of our main purposes in this work to show that the one loop potential might contain in itself the answer to this problem.
First we will use the nice feature present at one loop that the gauge, fermion and ghost fields do not interfere
with one another.
Then we will separate in Eq (\ref{not876}) the functional $\Delta^j$ which contains explicitly the spin interaction and K defined as,
\begin{equation}
K=-\partial^2+\Delta^1+\Delta^2
\label{def43567}
\end{equation}
which contains the rest.

We will use two methods to prove our point which stem from rewriting Eq(\ref{somformula64789}) as,
\begin{eqnarray}
&&\log \det\bigtriangleup_{r,j}=\log\det[-\partial^2+\bigtriangleup^1+\bigtriangleup^2+\bigtriangleup^j]
\nonumber\\
&&\log\det K +Tr [K^{-1}\bigtriangleup^j-\frac{(K^{-1}\bigtriangleup^j)^2}{2}+....]
\label{point1}
\end{eqnarray}

for the first method, and
\begin{eqnarray}
&&\log \det\bigtriangleup_{r,j}=\log\det[-\partial^2+\bigtriangleup^1+\bigtriangleup^2+\bigtriangleup^j]
\nonumber\\
&&\frac{1}{2}\log\det[K^2+(\bigtriangleup^j)^2]
\label{point2}
\end{eqnarray}

for the second one.

\subsection{First view}

Then  by introducing Eq (\ref{point1}) into Eq (\ref{eff435267}) we obtain the one loop potential as a function of $\Delta^j$:
\begin{eqnarray}
V=A_0+A_2(\bigtriangleup^j)^2+A_4(\bigtriangleup^j)^4+...
\label{func5634567}
\end{eqnarray}
What is the real significance of this equation written in terms of operators? First let us remember the general definition of the magnetization which is just the derivative of the vacuum energy(potential) with respect to the
magnetic field. Using the arguments given in appendix we can identify $F^{2n}=H^{2n}$ where H is the magnetic field. Retaining, as it is usual, only the first non-trivial term in Eq (\ref{func5634567})we obtain:
\begin{equation}
M_{paramag}=A_2\frac{\partial(\Delta^j)^2}{\partial H}=2A_2\frac{(\Delta^j)^2}{H^2}H
\label{def4356}
\end{equation}

Note that here $\Delta^j$ already contains one power of H. Then in terms of the paramagnetic magnetization,
\begin{eqnarray}
V=A_0+A_2(2A_2\Delta^j/H)^{-2}M^2_{paramag}+A_4(2A_2\Delta^j/H)^{-4}M^4_{paramag}+....
\label{def342567}
\end{eqnarray}

so there is a correspondence one to one between the powers of the magnetic field  and
the paramagnetic magnetization.
Since the higher order correction to this functional will be suppressed one expects that if a phase transition occurs this will be fairly well approximated by the first three terms. Thus the sign of the second term will be most relevant.
\begin{eqnarray}
A_2^{-1}\approx-[(\Delta^1)^2-2\Delta^2]=\frac{1}{12\pi^2}(N_f-N)\ln{\frac{q}{\Lambda}}
\label{import546738}
\end{eqnarray}

Here the term proportional to N contains the gauge ($\approx 2N$) and ghost ($\approx-N$) contributions.

\subsection{Second view}
We will use the standard functional approach for treating phase transition.
Thus the action in a background color field is written schematically as:
\begin{eqnarray}
\int  {\cal D}{\cal A}{\cal D}\Psi {\cal D}c \exp
[i\int  dx^4[-\frac{1}{4g^2}(F^a_{\mu\nu})^2+{\cal L}_{c.t}+\frac{N_f}{2}{\bar \Psi} \Delta_{r,1/2}\Psi+
-\frac{1}{2}{\cal A}\Delta_{g,1}{\cal A}+c\Delta_{G,0}c]]=
\nonumber\\
\exp[i\int d^4x(-\frac{1}{4g^2}(F_{\mu\nu}^2+{\cal L}_{c.t}]
(\det{\Delta_{G,1}})^{-\frac{1}{2}}(\det{\Delta_{r,1/2}})^{N_f/2}(\det{\Delta_{G,0}})^{+1}
\label{form75894}
\end{eqnarray}

Now  by using Eq (\ref{point2}) and retracing the steps we can write Eq (\ref{form75894}) as:
\begin{eqnarray}
&&\int {\cal D}{\cal A}{\cal D}\Psi {\cal D} c
\exp[i\int d^4x (-\frac{1}{4g^2}(F^a_{\mu\nu})^2
\nonumber\\
&&+{\cal L}_{c.t.}+
\frac{N_f}{4}{\bar \Psi}(K^2+(\Delta^j)^2)_{r,1/2}\Psi -\frac{1}{4}{\cal A}(K^2+(\Delta^j)^2)_{g,1}{\cal A}+
\frac{1}{2}c(K^2+(\Delta^j)^2)_{G,0}c]
\label{new546378}
\end{eqnarray}

Note that the formula contains gauge, scalar and fermion degrees of freedom. According to the standard procedure we separate the part of the action independent of spin into a functional $S_0$ and keep in $S_{int}$ only the spin dependent one. Then using Hubbard Stratonovich transformation we decouple $S_0$ and introduce the field M \cite{Belitz}. Then:
\begin{eqnarray}
&&Z=\int{\cal D}{\cal A}{\cal D}\Psi{\cal D}c e^{S_0+S_{int}}=
\nonumber\\
&&{\rm const}\times\int {\cal D}M e^{-g^2\int dx M^2}\langle e^{-2g^2\int dx M\Delta^j}\rangle_0
=
\nonumber\\
&&{\rm const}\times \int {\cal D}Me^{-\Phi(M)}
\label{func56478}
\end{eqnarray}

where:
\begin{eqnarray}
&&\Phi(M)=g^2\int dxM^2(x)-\ln\langle e^{-2g^2\int dx M\Delta^j}\rangle_0
=
\nonumber\\
&&\frac{1}{2}\int dx_1 dx_2 M(x_1)[\frac{\delta(x_1-x_2)}{g^2}-j^{(2)}(x_1-x_2)]M(x_2)+\frac{1}{3!}
\int dx_1 dx_2 dx_3 j^{(3)}(x_1,x_2,x_3)M(x_1)M(x_2)M(x_3)+...
\label{mag534627}
\end{eqnarray}

Here the j's are the analogs of the spin correlation functions and correspond practically to:
\begin{eqnarray}
&&j^{(2)}=(\Delta^j)^2/(F^2)=\frac{1}{8\pi^2}(4N-N_f)\ln{\frac{q}{\Lambda}}
\nonumber\\
&&j^{(3)}=(\Delta^j)^3/(F^3)...
\label{thej67456}
\end{eqnarray}

Since the term $\Delta^j$ is actually symmetric we will have only even contributions.

Now Eq (\ref{mag534627}) does not say much by itself. Let us study it close to the infrared fixed point of the theory \cite{collective}:
\begin{eqnarray}
\frac{1}{\alpha}=\frac{1}{6\pi}(11N-2N_f)\ln{\frac{q}{\Lambda}}+\frac{1}{\alpha^*}\ln{\frac{\alpha}{\alpha^*-\alpha}}
\label{per54678}
\end{eqnarray}
We used a perturbative approach thus we will consider the expression of the coupling constant in the perturbative  regime($q\gg\Lambda$):
\begin{equation}
\alpha\cong\frac{1}{\frac{1}{6\pi}(11N-2N_f)\ln{\frac{q}{\Lambda}}}
\label{app9078}
\end{equation}
and then:
\begin{equation}
\frac{1}{g^2}=\frac{1}{24\pi^2}(11N-2N_f)\ln{\frac{q}{\Lambda}}
\label{cp897}
\end{equation}

Substituting (\ref{app9078}) in Eq(\ref{mag534627}) we obtain :
\begin{eqnarray}
\Phi(M)=a_2M^2+a_4M^4+...
\label{coefsr54678}
\end{eqnarray}

where $a_4$ changes sign for large values of $N_f$ of no interest here and :
\begin{equation}
a_2=\frac{1}{24\pi^2}(-N+N_f)\ln{\frac{q}{\Lambda}}
\label{co674567}
\end{equation}

Thus we expect that if there is a phase transition this will occur at $N=N_f$.

\section{Diamagnetism}

Now one can consider the diamagnetic part in the expansion and repeat the calculations to obtain exactly the same results where this time the coefficients in part A are proportional to $(\Delta^j)^2$ and thus,
\begin{eqnarray}
V=A_0'+A_2'K^2+A_4'K^4+...
\label{pot6758}
\end{eqnarray}

where
\begin{equation}
(A_2')^{-1}=\frac{1}{8\pi^2}(N_f-4N)\ln{\frac{q}{\Lambda}}
\label{coeff786}
\end{equation}

The r.h.s. of Eq (\ref{coeff786}) being spin dependent does not contain any ghost contribution.

In part B the correlation  functions will be respectively
\begin{eqnarray}
&&k^{(2)}=K^2/(F^2)
\nonumber\\
&&k^{(4)}=K^4/(F^4)...
\label{newco657849}
\end{eqnarray}
so a similar result can be derived for the second case:
\begin{equation}
\Phi(M')=b_2M'^2+b_4M'^4+...
\label{sec345267}
\end{equation}

Here:
\begin{equation}
b_2=\frac{1}{8\pi^2}(N_f-4N)\ln{\frac{q}{\Lambda}}
\label{new45367}
\end{equation}

 Since the coefficient $b_4$ changes sign for a value of $N_f$ outside the range of interest a possible phase transition occurs at $N_f=4N$. We know from study of the gap equation that it is very possible that the chiral symmetry restoration takes place at $N_f\cong 4N$.

\section{Discussion}

The present work is based entirely on the Landau theory of phase transition. Thus in order to describe ferromagnets Landau uses magnetization as the order parameter and a Gibbs free energy of the form:
\begin{equation}
G(M)=A_0(T)+A_2(T)M^2+A_4(T)M^4+..
\label{gibbs}
\end{equation}
 Things like spontaneous symmetry breaking in electroweak theory find their meaning in the theory of superconductivity. We believe this language is fairly appropriate for things like chiral symmetry breaking too.
 Had we written the one loop effective potential in terms of the full magnetization we would have obtained transition between the confinement and abelian phase. But as we have shown the full potential can be written
 consistently in terms of the paramagnetic and diamagnetic operators. Of course we do not associate in normal settings paramagnetism an diamagnetism with different phases. A quantum field theory however is complicated enough for admitting various descriptions.
 On the other hand the secular approach requires the anomalous dimension of the fermion mass operator and at least the two loop potential. This information is obtained also in perturbation theory and contained fully after all in the tree level
 lagrangian as it is the case for our approach. If there is any connection between the two besides the obvious results it remains to be seen.

In \cite{Schechter} the authors proposed a description of the chiral symmetry phase transition based on instanton potential derived from its supersymmetric counterpart introduced by Seiberg in \cite{Seiberg}. There for example one needed two distinct forms of the potential for the case $N_f<N$ and $N_f>N$. The case $N=N_f$, where a singularity of the potential occurs is intriguing.

In the present work we obtained that something similar to a phase transition occurs both for $N=N_f$ and for $N_f=4N$.
While the latter can be easily related to chiral symmetry restoration which in most studies takes place at approximately $N_f=3.9N$ the first one is more problematic.
In the supersymmetric case Seiberg argued that $N=N_f$ corresponds to the appearance of new massless degrees of freedom associated with the baryons. A clue that something happen at $N=N_f$ can be the simple fact that one needs two different potential for the two regimes $N_f<N$ and $N_f>N$. What is then the symmetry which gets restored at $N=N_f$ is a question which answer we can only guess.
For example the possible symmetries for low energy QCD are:
\begin{eqnarray}
U(1)_A \times U(1)_B \times SU(3)_L \times SU(3)_R
\label{sym546738}
\end{eqnarray}

We know that at $N=N_f$ the chiral symmetry and the $U(1)_A$ should be broken. Thus if we eliminate this option the only possibility for a phase transition lies in $U(1)_B$. Something similar might arise for an arbitrary number of flavors and colors. However in order to be sure one needs explicit potential to describe it as in \cite{Schechter} or \cite{Seiberg}.

 We considered the well known one loop effective potential for $SU(N)$ gauge theories in a background field as function of two different quantities associated at the classical level with the diamagnetic and paramagnetic magnetizations. It is not necessary to be so but the potential displays two minima which can indicate phase transitions, one at $N_f=4N$ and the other one at $N_f=N$. The first one can be associated with chiral phase transition while the second one is arguably related to $U(1)_B$ breaking.

\section*{Acknowledgments} \vskip -.5cm
We are happy to thank R. Escribano, A. Fariborz and J. Schechter for support and encouragement and J. Schechter for useful comments on the manuscript.
This work has been supported by CICYT-FEDER-FPA2008-01430.

\appendix
\section{}

First let us mention that from the point of view of the present work the background field can be chosen as desired
since it plays no fundamental role in our derivation. Thus we can consider a color field which has all the components equal to zero except one for which we do not specify the spatial direction.
First let us analyze the terms involving an odd number of gluon fields coming form $\Delta^1$ and $\Delta^2$. Thus for the $N^2-1$(adjoint) representation,
\begin{eqnarray}
Tr({T^{a_1}T^{a_2}...T^{a_{2k+1}}})=-Tr({T^{a_{2k+1}}.... T^{a_1}})=0
\label{firstfunction}
\end{eqnarray}

by the antisymmetry of the corresponding matrices. Note that here we did not make any particular choice of the gauge fields.
For the fermion representation it is sufficient to quote that among matrices there will be antisymmetric ones by  a similar choice to that in the $SU(3)$ case. The representation will contain $N^2-1$ matrices where $N-1$ will be diagonal and the others are chosen according to the prescription $D_{ij}$ and $D_{ij}'$ have all the elements zero except
${ij}$ and ${ji}$ which are both 1 for $D_{ij}$, and one i and the other one -i for $D_{ij}'$. Thus we consider the field corresponding to one of the $D_{ij}'$'s different than zero an the other one equal to zero. By the virtue of the same argument as in Eq(\ref{firstfunction}) the odd terms will give zero.

It remains for us to consider the terms involving the gauge field tensor. Mixed terms will contain either an odd number of the generators who's trace is zero by the previous argument either an even number of the generators and an odd number of gauge tensors. Now the space time structure of the gauge tensor is determined entirely by the spin structure whatever this might be. Thus we expect,
\begin{eqnarray}
 F^{a1}_{\mu_1\mu_2}F^{a_2}_{\mu_2mu_3}... F^{a_{2k+1}}_{\mu_{2k+1}\mu_1}=0
 \label{tens4356}
\end{eqnarray}

by the same argument of the antisymmetry given previously.
So it is clear that we can choose the background magnetic field conveniently such that the odd powers of the gauge tensor are all equal to zero.

%Our method uses a perturbative approach; our hope that it as a physical significance is enforced by the fact that the %two functional that we study are the only ones that one can form consistently. For example one might be inclined to %choose the fermion spin function or the full fermion functional(to decouple the fermions)  but this leads to no %interesting physics.


\begin{thebibliography}{15}
\bibitem{Chivukula} R. S. Chivukula, Phys Rev D \textbf{55}, 5238 (1997).
\bibitem{Braun} J. Braun and H. Gies, JHEP 0606:024 (2006); J. Braun and H. Gies, arXiv:0912.4168 (2009);
H. Gies and J. Jaeckel, Eur. Phys. J. C46:433-438 (2006).
\bibitem{Klevansky} S. P. Klevansky, Rev. Mod. Phys. Vol 64, 649 (1992).  This is a thorough review of chiral
symmetry breaking (restoration) in the Nambu-Jona-Lasinio model.
\bibitem{collective}R. D. Pisarski and F Wilczek, Phys Rev D \textbf{29}, 338 (1983); A. Cohen and H. Georgi, Nucl Phys B 314(1984)7-24;
    T Apelquist, J. Terning and L. C. R Wijewardhana, arXiv:hep-ph/9602385; T. Apelquist, A. Ratnaweera, J. Terning and L. C. R. Wijewardhana, arXiv:hep-ph/9806472.
\bibitem{Schechter}F. Sannino and J. Schechter, Phys. Rev. D \textbf{57}, 170(1998); S. D. Hsu, F. Sannino and J. Schechter, Phys. Lett. B \textbf{427}, 300 (1998); F. Sannino and J. Schechter, arXiv:hep-ph/9903359 (1999).
\bibitem{Sannino} F. Sannino, arXiv:0911.0931 (2009).
\bibitem{Georgi} A. Cohen and H. Georgi, Nucl Phys. B \textbf{314}, 7-24(1989).
\bibitem{Nielsen}N. K. Nielsen, Am. J. Phys 49, 1171, 1981.
\bibitem{Vainshtein} A. Vainshtein, V. Zakharov, V. Novikov and M. Shifman, Sov. Phys. Usp. {\bf 25}, 195, 1982.
\bibitem{Peskin} M. E. Peskin and D.V. Schroeder, " Quantum Field Theory", Perseus Books Publishing, L. L. C., 1995(pg
533-543).
\bibitem{Belitz} D. Belitz, T. R. Kirkpatrick, T. Vojta, arXiv:cond-mat/0109547.
\bibitem{Seiberg} N. Seiberg, Phys. Rev D 49, 6857 (1994); N. Seiberg, Nucl Phys. B 435, 129 (1995).

\end{thebibliography}
\end{document}